\documentclass[conference]{IEEEtran}
\IEEEoverridecommandlockouts
\usepackage{amsmath,amssymb,amsfonts}
\usepackage{acronym}
\usepackage{algorithmic}
\usepackage{graphicx}
\usepackage{textcomp}
\usepackage{xcolor}
\usepackage{tikz}
\usepackage[a4paper, total={184mm,239mm}]{geometry}
\def\BibTeX{{\rm B\kern-.05em{\sc i\kern-.025em b}\kern-.08em
    T\kern-.1667em\lower.7ex\hbox{E}\kern-.125emX}}
\usepackage[numbers]{natbib}
\usepackage{multirow}
\usepackage{threeparttable}
\usepackage{subfigure}
\usepackage{hyperref}
\DeclareRobustCommand*{\IEEEauthorrefmark}[1]{%
  \raisebox{0pt}[0pt][0pt]{\textsuperscript{\footnotesize\ensuremath{#1}}}}
  
\begin{document}

\newcommand\copyrighttext{%
  \footnotesize \textcopyright 2026 IEEE. Personal use of this material is permitted.
  Permission from IEEE must be obtained for all other uses, in any current or future
  media, including reprinting/republishing this material for advertising or promotional
  purposes, creating new collective works, for resale or redistribution to servers or
  lists, or reuse of any copyrighted component of this work in other works.}
\newcommand\copyrightnotice{%
\begin{tikzpicture}[remember picture,overlay]
\node[anchor=south,yshift=10pt] at (current page.south) {\fbox{\parbox{\dimexpr\textwidth-\fboxsep-\fboxrule\relax}{\copyrighttext}}};
\end{tikzpicture}%
}

\acrodef{CPU}[CPU]{central processing unit}
\acrodef{FPGA}[FPGA]{field-programmable gate array}
\acrodef{GPU}[GPU]{graphics processing unit}
\acrodef{PSF}[PSF]{point-spread function}
\acrodef{NISQ}[NISQ]{noisy intermediate-scale quantum}
\acrodef{PL}[PL]{programmable logic}
\acrodef{PS}[PS]{processing system}
\acrodef{DRAM}[DRAM]{Dynamic Random Access Memory}
\acrodef{DDR}[DDR]{Double Data Rate}
\acrodef{IP}[IP]{Intellectual Property}
\acrodef{MMIO}[MMIO]{Memory-Mapped I/O}
\acrodef{HLS}[HLS]{High-level Synthesis}
\acrodef{std}[std]{standard deviation}
\acrodef{HPC}[HPC]{high performance computing}
\acrodef{API}[API]{application programming interface }
\acrodef{NAQC}[NAQC]{neutral atom quantum computer}
\acrodef{NA}[NA]{neutral atom}
\acrodef{LUT}[LUT]{lookup table}
\acrodef{FF}[FF]{flip-flop}
\acrodef{DSP}[DSP]{digital signal processing}
\acrodef{BRAM}[BRAM]{block RAM}

\title{
Efficient Image Reconstruction Architecture for Neutral Atom Quantum Computing
\thanks{\IEEEauthorrefmark{*} These authors contributed equally to this work.

This work was funded by the German Federal Ministry of Education and Research (BMBF) under the funding program \textit{Quantum Technologies - From Basic Research to Market} under contract numbers 13N16077 and 13N16087, as well as from the Munich Quantum Valley~(MQV), which is supported by the Bavarian State Government with funds from the Hightech Agenda Bayern.}
}


\author{\IEEEauthorblockN{Jonas Winklmann\IEEEauthorrefmark{*,1}, Yian Yu\IEEEauthorrefmark{*,2}, Xiaorang Guo\IEEEauthorrefmark{1}, Korbinian Staudacher\IEEEauthorrefmark{2}, and Martin Schulz\IEEEauthorrefmark{1}}

\IEEEauthorblockA{\IEEEauthorrefmark{1}Chair of Computer Architecture and Parallel Systems, Technical University of Munich, Garching, Germany \\ 
\IEEEauthorrefmark{2}MNM-Team, Ludwig-Maximilians-Universität München, Munich, Germany\\
Email: {\{jonas.winklmann, xiaorang.guo, martin.w.j.schulz\}@tum.de, 
yian.yu@campus.lmu.de, staudacher@nm.ifi.lmu.de}}
}

\bibliographystyle{IEEEtranN}

\maketitle
\copyrightnotice
\begin{abstract}
In recent years, \acp{NAQC} have attracted a lot of attention, primarily due to their long coherence times and good scalability. 
One of their main drawbacks is their comparatively time-consuming control overhead, with one of the main contributing procedures being the detection of individual atoms and measurement of their states, each occurring at least once per compute cycle and requiring fluorescence imaging and subsequent image analysis. 

To reduce the required time budget, we propose a highly-parallel atom-detection accelerator for tweezer-based \acp{NAQC}. Building on an existing solution, our design combines algorithm-level optimization with a \ac{FPGA} implementation to maximize parallelism and reduce the run time of the image analysis process. 
Our design can analyze a 256$\times$256-pixel image representing a 10$\times$10 atom array in just 115 $\mu$s on a Xilinx UltraScale+ \ac{FPGA}. Compared to the original CPU baseline and our optimized CPU version, we achieve about 34.9$\times$ and 6.3$\times$ speedup of the reconstruction time, respectively. Moreover, this work also contributes to the ongoing efforts toward fully integrated \ac{FPGA}-based control systems for \acp{NAQC}.
\end{abstract}


\section{Introduction}
\label{section:intro}
Typical \acfp{NAQC} require at least two atom detection steps, one for initial detection and one for the final readout, with possibly more to detect atom losses during sorting or, especially for fault-tolerant computing~\cite{mid_FTQC}, for mid-circuit measurements. Most other required steps, such as gate execution and atom loading, are comparatively fast. As a result, the detection process ought to be performed rapidly, making parallel algorithms highly desirable. 

While the capability to resolve single atoms through microscopic imaging is well-explained and influential in many adjacent fields \cite{Schlosser:1, Bakr:1, Haller:1, Sherson:1, Morgado:1}, the ongoing transition of \ac{NAQC} from lab experiment to generally useful computational platform imposes new challenges in terms of robustness and independence. \Acf{FPGA}-based control architectures emerge as a promising solution that fulfills all these requirements, delivering a low-latency, programmable hardware required for real-time atom detection tasks.




Therefore, in this work, we propose a highly parallel image deconvolution solution for atom detection based on the aforementioned hardware devices. Our solution utilizes the hardware-software co-design strategy to optimize an existing atom detection method, namely the projection-based state-reconstruction algorithm presented in work~\cite{Wei:1}. Through this approach, we obtain a CPU-optimized version with high inherent parallelism and streamlined logic. Building on this, we develop an \ac{FPGA}-based reconstruction accelerator that, once connected to a camera, can directly bridge the gap between image generation and \ac{FPGA}-based rearrangement \cite{Guo:1,wang2023accelerating} or readout procedures. Overall, with many microwave pulse generation platforms already being based on \acp{FPGA}~\cite{stefanazzi2022qick,liu2025risc,xu2023qubic}, the vision of a fully integrated sorting and readout device is nearing completion.

\section{System Architecture}
\label{chapters:impl}
\label{sec:sys_overview}
\begin{figure}[tb] 
  \centering
  \includegraphics[width=0.75\linewidth]{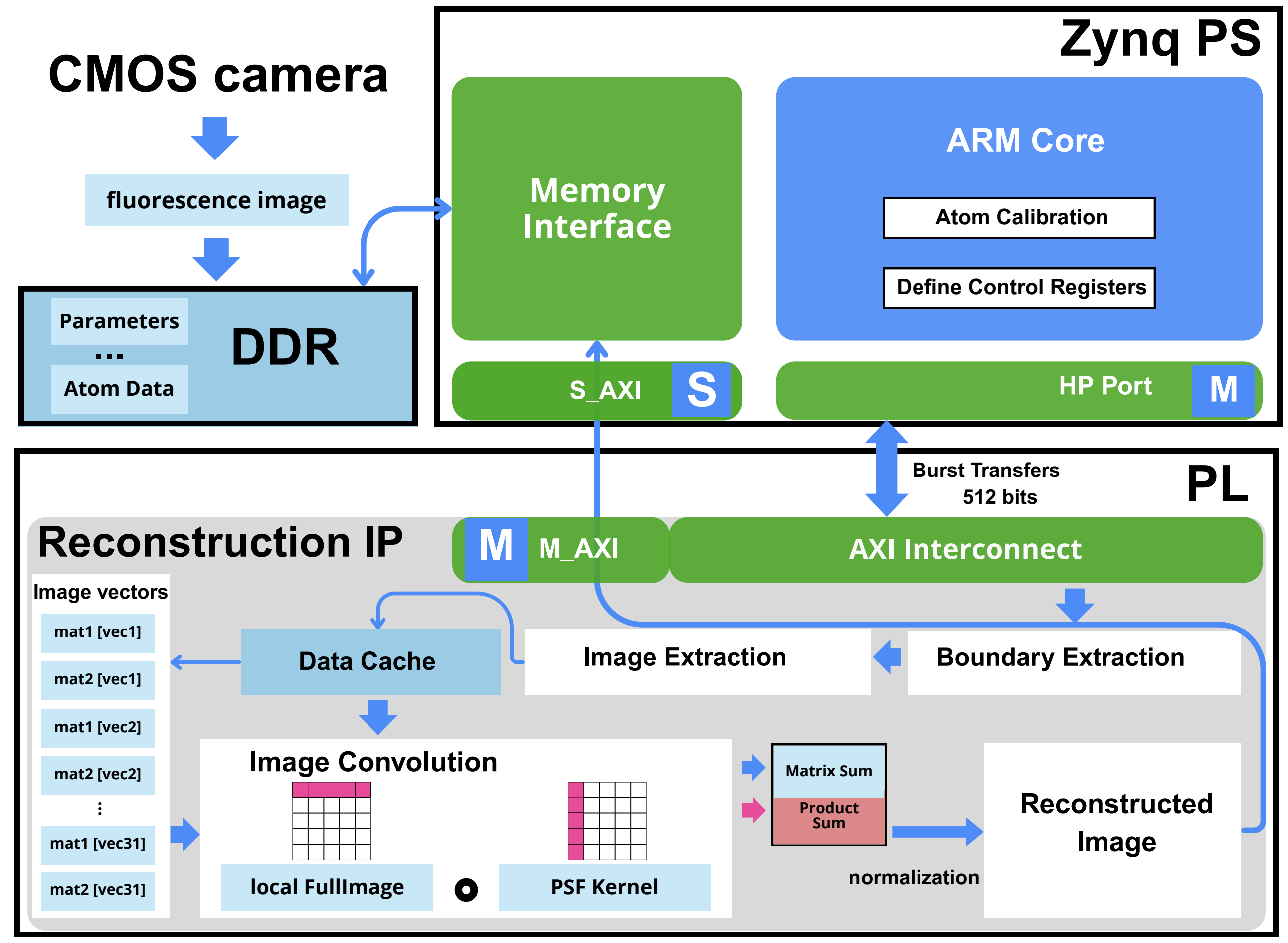}
  \caption{Overall Architecture of the Reconstruction Accelerator. In this figure, \textit{mat1} and \textit{mat2} represent the PSF kernel and the atom's image, respectively.}
  \label{fig:design}
\end{figure}

Fig.~\ref{fig:design} shows the overall system architecture of our FPGA-based accelerator, where the \ac{PL} and \ac{PS} work together to process atom positions and apply the state-reconstruction algorithm. The \ac{PS} side is mainly responsible for software-driven primitive atom calibration, control register configuration, and high-level data communication. Meanwhile, the \ac{PL} side implements dedicated hardware development, which provides substantial benefits in the parallel execution of the state-reconstruction process.

The entire reconstruction accelerator, a customized \ac{IP} core, adopts the dataflow design, as shown in the bottom part of Fig.~\ref{fig:design}, enabling task-level parallelism across different modules. Before execution, the DDR memory is initialized with the atom position grid, a $31 \times 31$ PSF kernel, and the fluorescence image. The \textit{Boundary Extraction} module first identifies the local region of interest for each atom. Following this, the \textit{Image Extraction} module fetches the corresponding pixel data and PSF kernel through a 512-bit AXI interface. This high-bandwidth data is then decoded to 32-bit precision, and the PSF kernel is pre-processed into a projector to facilitate the subsequent \textit{Image Convolution stage}.

Within the \textit{image convolution} module, two parallel computational paths are executed: (1) an element-wise matrix multiplication between the local image detail and the projector kernel to derive the \textit{product sum} (indicated by pink arrows in Fig.~\ref{fig:design}), and (2) a summation of all elements within the projector matrix to obtain the \textit{matrix sum} (indicated by blue arrows in Fig.~\ref{fig:design}). To achieve high throughput, the matrix operations are decomposed into 31 concurrent vector processing units, each featuring internal parallelization. Furthermore, a logarithmic reduction algorithm is implemented to compute vector sums via a four-stage adder tree. This architecture reduces the computational complexity from $\mathcal{O}(n)$ to $\mathcal{O}(\log n)$, enabling the summation of 31 elements to be completed within only five clock cycles.


Ultimately, the \textit{output aggregation} module calculates a normalized value for each convolution result from the previous step, thereby providing the normalized brightness at each atomic site.
\section{Experiments and Evaluation}
\label{chapter:eva}
We evaluate this work based on the quality of the reconstructed image, reconstruction run time, and FPGA resource utilization. For run time comparison, we benchmark the accelerator against the original CPU baseline~\cite{Wei:1} and our optimized CPU implementation (CPU-opt). The FPGA design is implemented on a Xilinx ZCU216 board~\cite{AMDrfsoc} at 100 MHz.

\begin{figure}[tb]
  \centering
  \subfigure[Raw image]{\includegraphics[width=0.28\linewidth]{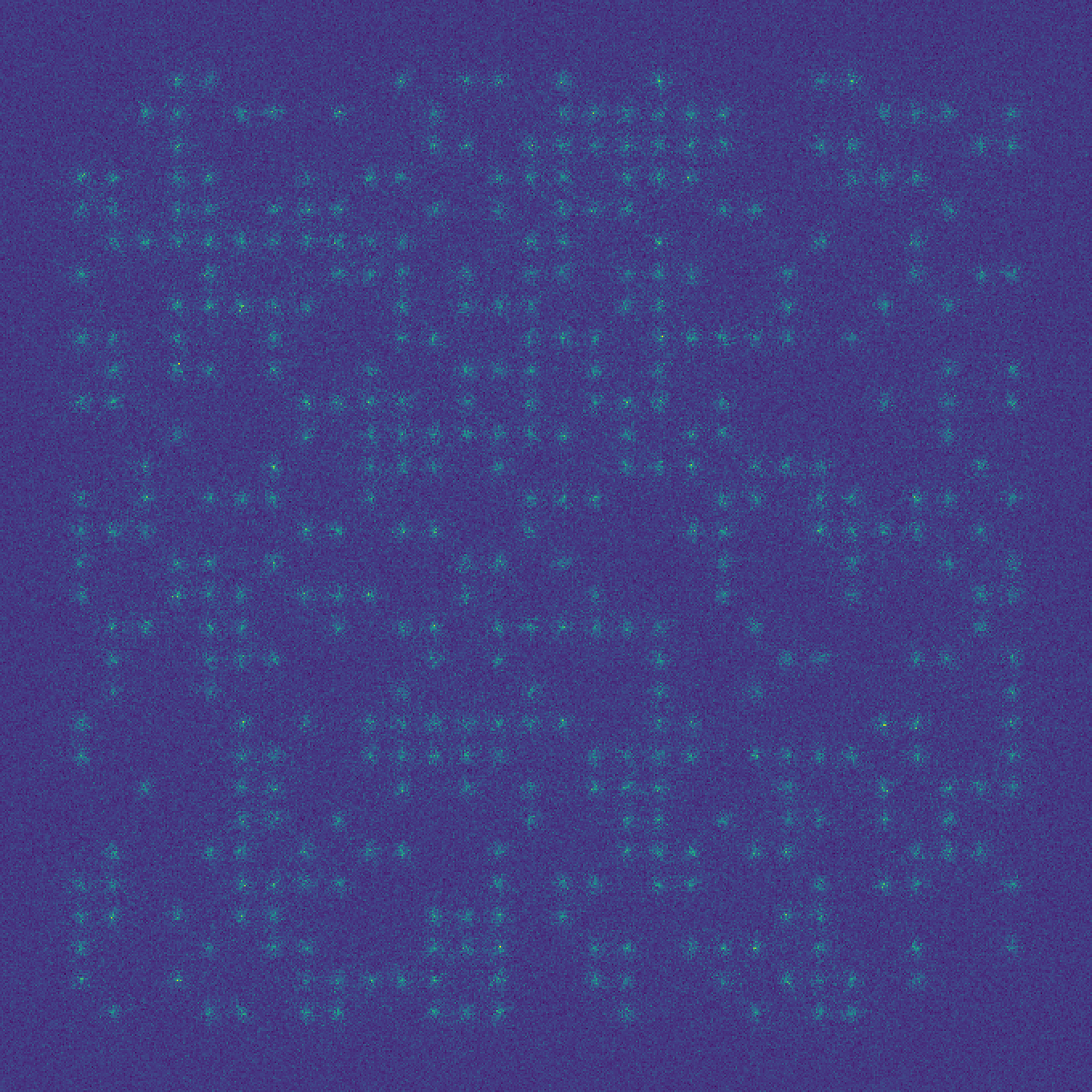}
  \label{fig:reconstruction_fig_camera}}
  \hfill
  \subfigure[Reconstructed image]{\includegraphics[width=0.28\linewidth,]{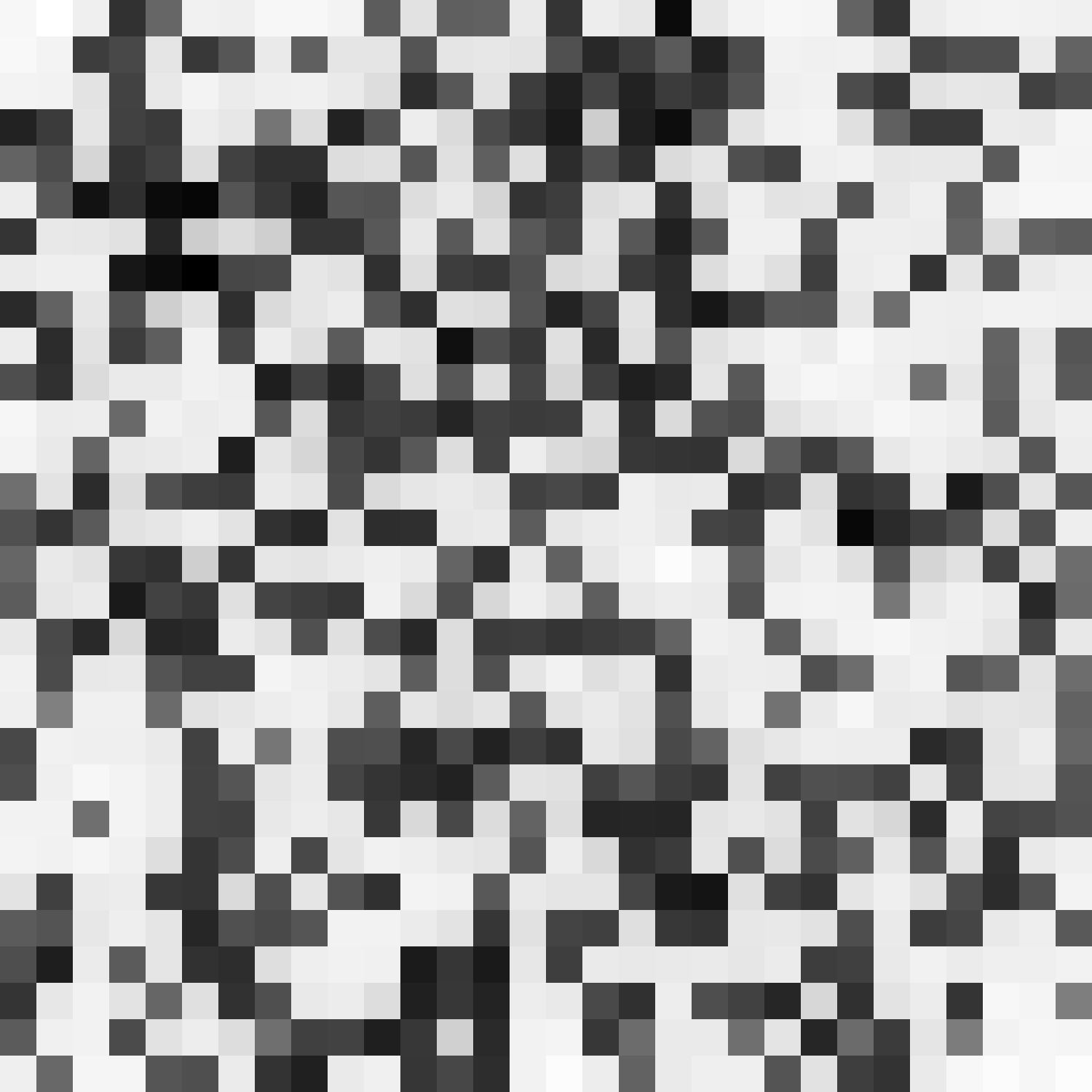}
  \label{fig:reconstruction_fig_recons}}
  \hfill
  \subfigure[Thresholded image]{\includegraphics[width=0.28\linewidth,]{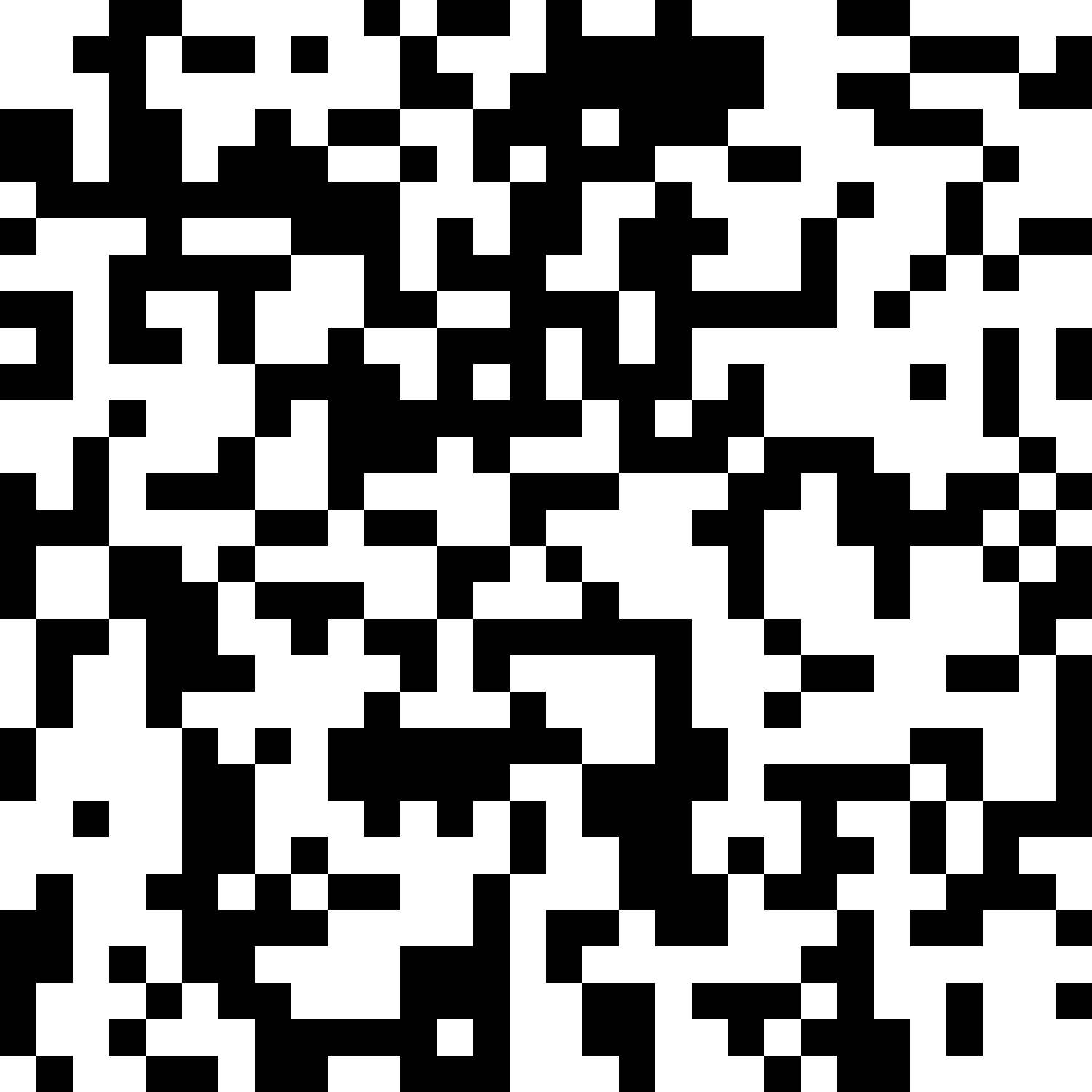}
  \label{fig:truth_fig_recons}}
  \caption{Result of the detection algorithm with a 30$\times$30 atom array as an example. (a) A raw image of the atom array captured by a camera (simulated). (b) The output of our reconstruction accelerator. Darker pixels denote higher reconstructed brightness. (c) Thresholded boolean result using a calibrated threshold. Black denotes a detected atom.}
  \label{fig:reconstruction_fig}
\end{figure}

\subsection{Reconstructed Image}
Fig.~\ref{fig:reconstruction_fig} shows the result of applying the reconstruction algorithm to a 30$\times$30 atom array. After the reconstruction procedure, we obtain a so-called \textit{emission matrix}, where its values of emissions indicate the brightness of atoms. Denoting higher emission values by darker pixels, we generate the reconstructed image of atoms shown in Fig.~\ref{fig:reconstruction_fig_recons}. The algorithm's precision has been shown previously~\cite{Winklmann:1}, so ensuring that our adaptation performs equivalently to the original is sufficient. This is the case up to insignificant differences that can be attributed to rounding errors.


\subsection{Run Time of Reconstruction}

\begin{figure}[tb]
\centering
{\includegraphics[width=0.6\linewidth]{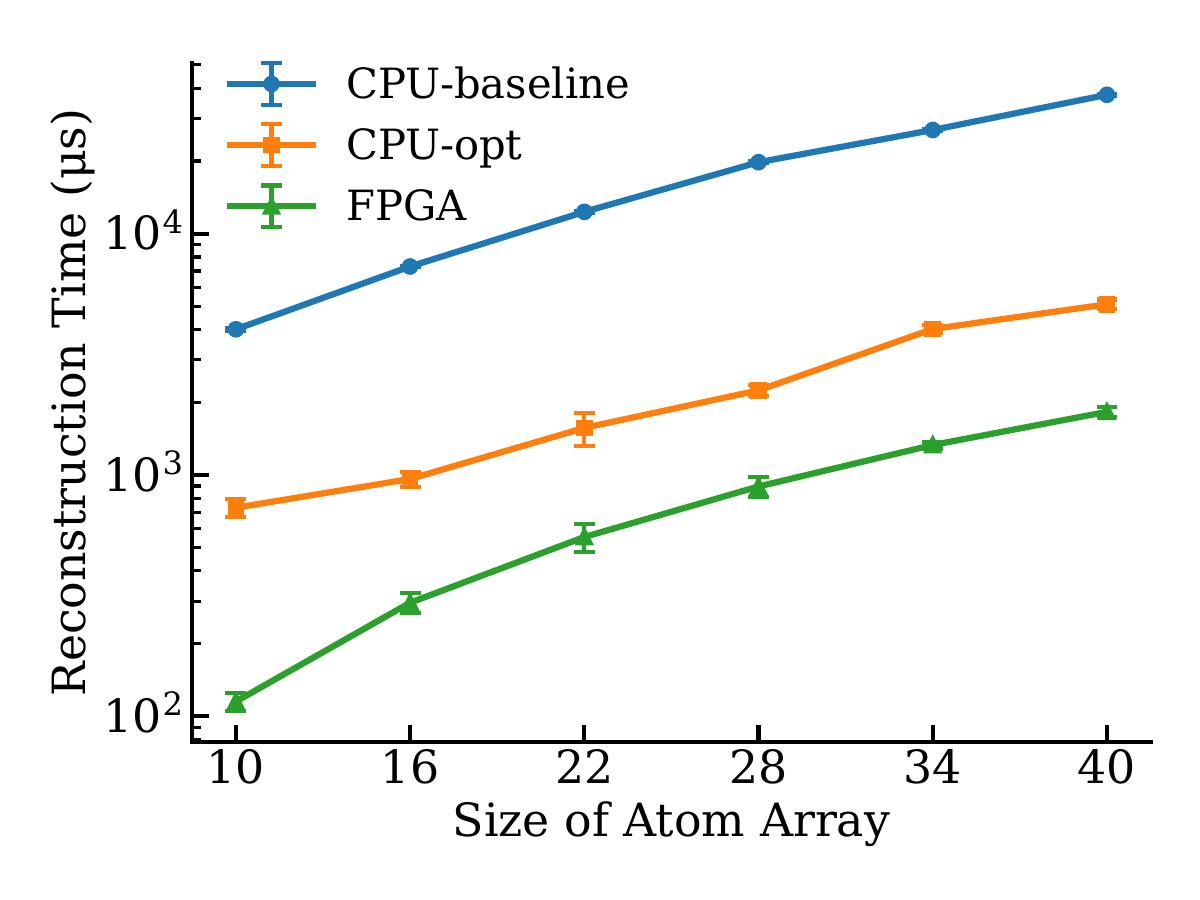}
\caption{Run time comparison of the reconstruction process for various sizes of the atom array (10$\times$10 to 40$\times$40) among CPU-baseline, CPU-opt, and FPGA. Error bars represent the standard deviations of the run time.}
\label{fig:reconstruction_res}
}
\end{figure}

As illustrated in Fig.~\ref{fig:reconstruction_res}, we compare the reconstruction run time for various array sizes of the original algorithm (CPU-baseline), the CPU-optimized version (CPU-opt), and the proposed FPGA accelerator. The Y-axis is plotted on a logarithmic scale, with error bars representing the standard deviation of run time across 50 independent trials for each data point. The results demonstrate that our FPGA accelerator not only consistently outperforms the CPU counterparts in terms of speed but also exhibits superior stability with minimal run-time variance. Specifically, for a $10 \times 10$ array, the FPGA achieves a speedup of 34.9$\times$ ($115~\mu$s vs. $4012~\mu$s) over the CPU-baseline and 6.3$\times$ ($115~\mu$s vs. $730~\mu$s) over the CPU-opt.

\subsection{Resource Utilization}
Resource utilization is a critical metric for evaluating scalability. Thanks to our fixed parallelization and time-multiplexing design, resource consumption remains constant, independent of the atom array size. Based on the report of the Vivado implementation, the accelerator is highly efficient, utilizing only one-quarter of \acp{LUT} and $\sim$15\% of \acp{FF}, with negligible \ac{DSP} and \ac{BRAM} usage. This compact footprint reserves significant capacity for additional logic, making the design well-suited for integration into unified quantum control systems and future HPC-quantum platforms (HPCQC)~\cite{UQP,dobler2025survey,ramsauer2025towards}.

\section{Conclusion}
\label{chapter:con}

This work presented a highly parallel atom-detection accelerator for atom detection in \acp{NAQC}. By employing a software-hardware co-design strategy, we leverage algorithm-level optimization and an FPGA-based parallel architecture to minimize reconstruction latency. Our implementation achieves an ultra-low latency of 115 $\mu$s for a 10$\times$10 array and 1.825 ms for a 40$\times$40 array, representing a 34.9$\times$ speedup over the CPU baseline and 6.3$\times$ over an optimized CPU version. The design is characterized by its low hardware cost and excellent scalability, demonstrating significant potential for integration into \acp{NAQC} control systems.

\bibliography{bibliography.bib}

\end{document}